# ntLink: a toolkit for *de novo* genome assembly scaffolding and mapping using long reads

Lauren Coombe[1]* ([lcoombe@bcgsc.ca](lcoombe@bcgsc.ca))

René L. Warren[1] ([rwarren@bcgsc.ca](rwarren@bcgsc.ca))

Johnathan Wong[1] ([jowong@bcgsc.ca](jowong@bcgsc.ca))

Vladimir Nikolic[1] ([vnikolic@bcgsc.ca](vnikolic@bcgsc.ca))

Inanc Birol[1]* ([ibirol@bcgsc.ca](ibirol@bcgsc.ca))

[1]Canada's Michael Smith Genome Sciences Centre at BC Cancer, 570 W 7th Ave, Vancouver, BC V5Z 4S6, 604-707-5900

*Corresponding authors

## Abstract

With the increasing affordability and accessibility of genome sequencing data, *de novo* genome assembly is an important first step to a wide variety of downstream studies and analyses. Therefore, bioinformatics tools that enable the generation of high-quality genome assemblies in a computationally efficient manner are essential. Recent developments in long-read sequencing technologies have greatly benefited genome assembly work, including scaffolding, by providing long-range evidence that can aid in resolving the challenging repetitive regions of complex genomes. ntLink is a flexible and resource-efficient genome scaffolding tool that utilizes long-read sequencing data to improve upon draft genome assemblies built from any sequencing technologies, including the same long reads. Instead of using read alignments to identify candidate joins, ntLink utilizes minimizer-based mappings to infer how input sequences should be ordered and oriented into scaffolds. Recent improvements to ntLink have added important features such as overlap detection, gap-filling and in-code scaffolding iterations. Here, we present three basic protocols demonstrating how to use each of these new features to yield highly contiguous genome assemblies, while still maintaining ntLink's proven computational efficiency. Further, as we illustrate in the alternate protocols, the lightweight minimizer-based mappings that enable ntLink scaffolding can also be utilized for other downstream applications, such as misassembly detection. With its modularity and multiple modes of execution, ntLink has broad benefit to the genomics community, from genome scaffolding and beyond. ntLink is an open-source project and is freely available from [https://github.com/bcgsc/ntLink](https://github.com/bcgsc/ntLink).

Basic Protocol 1: ntLink scaffolding using overlap detection

Basic Protocol 2: ntLink scaffolding with gap-filling

Basic Protocol 3: Running in-code iterations of ntLink scaffolding

Alternate Protocol 1: Generating long-read to contig mappings with ntLink

Alternate Protocol 2: Using ntLink mappings for genome assembly correction with Tigmint-long

Support Protocol: Installing ntLink

## Keywords:

## Introduction

Generating high-quality *de novo* genome assemblies for both model and non-model organisms opens the door to a plethora of important downstream studies, including annotation, structural variant analysis and population studies, to name a few (Logsdon et al., 2020). In recent years, long-read genome sequencing technology from Oxford Nanopore Technologies PLC (ONT, Oxford,

UK) and Pacific Biosciences of California, Inc. (PacBio, Menlo Park, CA) have gained in popularity. Although long reads still have a higher base error rate than typical short read sequencing technologies, such as those from Illumina, their lengths can range from kilobases to megabases, orders of magnitude longer than the typical read length of 150-300 bp for short reads. This length distribution enables the long reads to span over the numerous repetitive elements present in complex genomes, therefore allowing for repeats to be resolved in a draft assembly. With the continued improvement and accessibility of long-read genome sequencing, there is great opportunity to harness the rich information of this data type for facilitating and improving *de novo* genome assemblies.

While current state-of-the-art *de novo* long-read assemblers such as Flye (Kolmogorov et al., 2019) are generating highly contiguous assemblies, we observe that they still do not fully exhaust or necessarily correctly use the long-range evidence inherent in long reads (Coombe et al., 2021). Therefore, there is great value in stand-alone genome assembly scaffolders, such as LINKS (Warren et al., 2015), OPERA-LG (Gao et al., 2016), LRScaf (Qin et al., 2019), and our alignment-free scaffolder ntLink (Coombe et al., 2021). ntLink is a lightweight, minimizer-based long-read scaffolding tool that was previously published as a central step in the correction and scaffolding pipeline LongStitch (Coombe et al., 2021). More recently, ntLink was integrated as a key step in our *de novo* long-read assembler GoldRush (Wong et al., 2022). ntLink uses long-read evidence to further contiguate draft assemblies from any sequencing technology. Instead of using alignments of long reads to the draft assembly, like many state-of-the-art long-read scaffolding tools, ntLink uses minimizer mappings (leveraging particular subsets or sketches of the sequence *k*-mers), which we have shown to be effective for assembly scaffolding as it confers considerable computational benefit.

ntLink is a flexible toolkit which can be run in various modes depending on the desired user output (Figure 1), with multiple new functionalities introduced since the published version. For each basic protocol, the input files provided by the users include the long reads and a draft assembly to be improved, and the main output file is a scaffolded assembly in FASTA format. The core functionality of ntLink uses the long-read evidence and generated minimizers to infer how the input contigs (draft assembly sequences) should be ordered and oriented, and subsequently performs these joins. The basic protocols differ in additional steps that are performed to more accurately join contigs together, fill ambiguous sequences (i.e. gap-filling), and enhance the final contiguity of the assembly. In earlier versions of ntLink, contigs were joined together end-to-end naively, whether they overlapped in the genomic space or not. The overlap detection feature of ntLink identifies cases where adjacent contigs overlap, and trims them to remove the overlapping regions prior to concatenation (Basic Protocol 1, Figure 1A). Furthermore, scaffolding assemblies generally introduces gaps, or ambiguous nucleotide bases ("N"s), between joined contigs. The gap-filling feature of ntLink instead fills gaps with bases from a representative read supporting the join (Basic Protocol 2, Figure 1B). Finally, running additional iterations of ntLink can further improve the contiguity of the final output assembly. To take advantage of these contiguity gains efficiently, ntLink can run in-code scaffolding iterations (also termed rounds hereon) powered by a coordinate liftover module (Basic Protocol 3, Figure 1C).

Although ntLink was developed as a scaffolding tool, the initial minimizer-guided step of mapping long reads to the draft assembly can also benefit other bioinformatics utilities that require approximate mapping information, including the ntEdit+Sealer polishing component of the GoldRush assembler (J. X. Li et al., 2022; Paulino et al., 2015; Warren et al., 2019; Wong et al., 2022) and Tigmint-long (Coombe et al., 2021; Jackman et al., 2018), an assembly correction tool also integrated in the LongStitch pipeline. In Alternate Protocols 1 and 2, we highlight the usage of ntLink as a mapping tool, and, as an example of using ntLink mappings in a different downstream application, showcase how these mappings can be utilized in Tigmint-long.

With its numerous modes and available protocols, ntLink is a flexible and wide-reaching tool for improving *de novo* genome assemblies and helping researchers better leverage their long-read sequencing data.

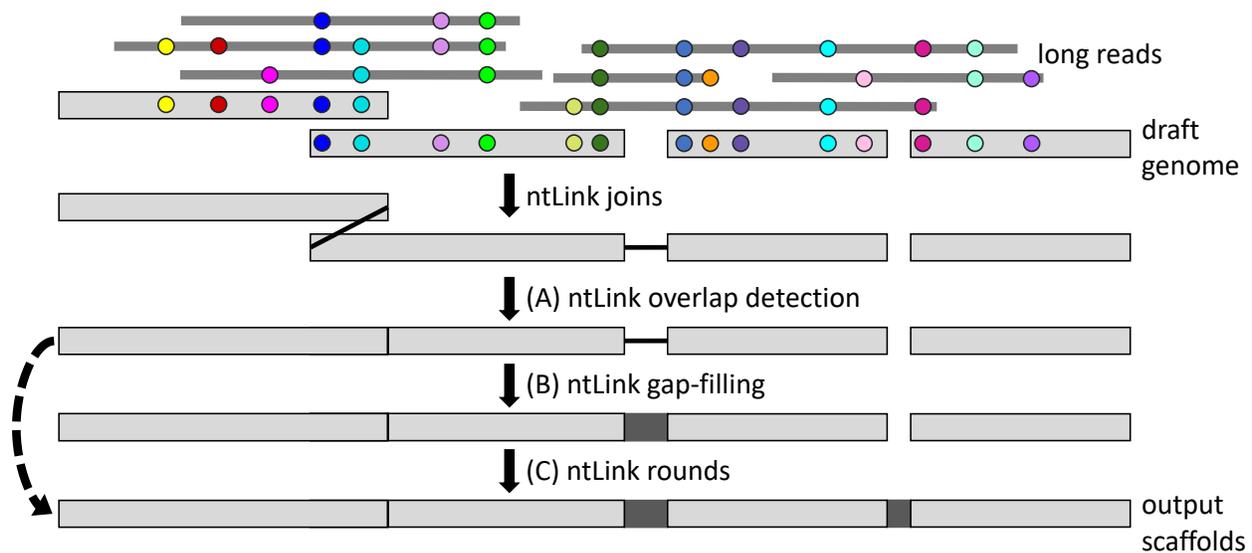

**Figure 1.** Flowchart showing the various available features for ntLink scaffolding. ntLink uses minimizers (indicated by coloured circles) to map the input long reads to the draft genome. Identical minimizers are represented by the same colour. ntLink uses these mappings to infer how the input contigs (draft genome sequences) should be ordered and oriented to produce genome scaffolds. (A) ntLink can also use minimizers to detect when adjacent contigs have an overlapping region, and resolves these overlaps (indicated by the vertical black line), as described in Basic Protocol 1. (B) In addition, as demonstrated in Basic Protocol 2, ntLink can use the input long-read information and minimizers to fill gaps between joined contigs (dark grey box). (C) Finally, running multiple in-code rounds, or iterations, of ntLink scaffolding, as shown in Basic Protocol 3, can maximize the contiguity of the final output scaffolds. The ntLink rounds can be run with or without gap-filling, as indicated by the dashed transitive arrow.

## Strategic Planning

**Hardware**

ntLink is a command-line tool, which can be run on 64-bit Linux or MacOS operating systems with sufficient available RAM (random-access memory). The amount of RAM and disk space required for running ntLink varies with the draft genome size and coverage of the long-read dataset. See Table 1 for the peak memory and disk space usage for representative ntLink scaffolding runs using four different species with varying genome sizes. The wall-clock time for each ntLink run is also included in Table 1. For the basic and alternate protocols, each step uses 5 threads, but this can be adjusted depending on the specifications of the user's machine. The parameter *t* controls the number of threads used (where applicable), and is shown in each corresponding command.

**Table 1**. Example peak memory usage, disk space usage and wall-clock times when running ntLink scaffolding on draft genomes of various sizes. The disk space usage of the input draft assembly and long reads files are not included in the benchmarks[a]. See Supplementary Tables 1 and 2 for more information about the data used for these example ntLink runs.

| Species | Approximate genome size (Mbp) | Fold coverage of long reads | Peak memory usage (GB) | Peak disk space usage (GB) | Wall-clock time (min) |
|---|---|---|---|---|---|
| *Caenorhabditis elegans* | 100 | 93 | 0.9 | 0.9 | 6.5 |
| *Oryza sativa* | 373 | 62 | 3.4 | 2.8 | 21.4 |
| *Solanum lycopersicum* | 824 | 72 | 5.4 | 5.6 | 43.4 |
| *Homo sapiens* | 3,055 | 50 | 20.8 | 25.0 | 136.0 (2h16m) |

[a]Default ntLink parameters were used for each example run, except for *Solanum lycopersicum*, which used k=64, w=250. One round of ntLink, including the gap-filling option, was run with each dataset.

**Software**

ntLink is available from the conda package manager for a more straightforward installation. Users can also install ntLink from the provided source code on GitHub. Detailed instructions for installing ntLink are available in the Support Protocol.

**Files**

Each ntLink protocol requires input long sequencing reads (in FASTA or FASTQ format), and an input draft genome assembly to be scaffolded (in FASTA format). Both files can be in single-line or multi-line (standard) FASTA format.

## Basic Protocol 1: ntLink scaffolding using overlap detection

Basic Protocol 1 describes running ntLink to scaffold an input draft assembly using long reads, with overlap detection enabled. The overlap detection functionality identifies when adjacent contigs (draft assembly sequences) overlap in genomic space, and trims the contigs to ensure the sequences are merged without duplicating this overlapping sequence.

ntLink leverages the long-range information inherent in long-read sequencing data to scaffold an input draft assembly. First, ntLink maps the long reads to a draft assembly using a minimizer-based approach. Long-read mappings that span multiple contigs provide evidence that these contigs should be joined together. These mappings are also used to estimate gap sizes between the contigs. After ntLink determines the sequences of oriented contigs to be joined together as scaffolds, the overlap detection feature identifies adjacent contigs that have a putative overlap (indicated by a negative estimated gap size), and resolves the overlaps. Finally, the ordered and oriented scaffolds are output in FASTA format.

### Necessary Resources:

*Hardware*

This protocol requires a 64-bit Linux or MacOS operating system with sufficient RAM and available disk space (See Strategic Planning for more information).

*Software*

The following software must be installed and available in your PATH environment variable:
- SRA toolkit (v3.0.0+): (*https://github.com/ncbi/sra-tools*)
- curl: (*https://curl.se/*)
- Python 3.7+: (*https://www.python.org/*)
- ntLink (v1.3.7+): (*https://github.com/bcgsc/ntLink*)
- ABySS (v2.3.0+): (*https://github.com/bcgsc/abyss*)
- QUAST (v5.2.0+): (*https://github.com/ablab/quast*)

*Files*

The input files for ntLink are long genome sequencing reads and a draft genome assembly. The long sequencing reads can be provided in FASTA or FASTQ format, either compressed with gzip or uncompressed. The input draft assembly to be scaffolded should be in FASTA format (multi-line or single-line).

*Sample Files*

To demonstrate the usage of ntLink in Basic Protocol 1, we will scaffold a *C. elegans* draft assembly with a corresponding *C. elegans* Oxford Nanopore long-read dataset. The *C. elegans* long reads are available from the Sequence Read Archive (SRA) under accession SRR10028109. The draft assembly is a Flye (Kolmogorov et al., 2019) assembly of the same *C. elegans* long reads, and is available from https://doi.org/10.5281/zenodo.7526395. To assess the genome assemblies generated from ntLink, a *C. elegans* N2 (Bristol strain) reference genome (accession GCA_000002985.3) will be used. There are detailed steps in the protocol to guide the user in downloading these files.

### Protocol steps:

1. Install ntLink
   *See Support Protocol for detailed instructions and options for installing ntLink.*

2. Install protocol-specific dependencies curl, SRA toolkit, and QUAST.
   *Option A: Use conda package manager*

i. If Option A of the Support Protocol was used to install ntLink, the protocol-specific dependencies can be installed in the same conda environment.

```
conda activate ntlink_env
conda install -y -c bioconda -c conda-forge curl quast 'sra-tools>=2.10.2'
```

*Option B: Install from source*
i. Install curl
   *Many servers will already have curl installed. To check if curl is available:*

   ```
   which curl
   ```

   *If you see the path to a curl installation, curl is already installed and you can continue to part (ii) to install QUAST. Otherwise, follow the next steps.*
   a. *Go to https://curl.se/download.html, and find the tarball for the latest released version. Version 7.86.0 is used below to illustrate the steps. Change your terminal's current directory to the location where you would like curl installed, download the tarball, extract the tarball and change your directory into the downloaded curl directory.*

   ```
   cd /path/to/new/curl/installation
   wget https://curl.se/download/curl-7.86.0.tar.gz
   tar zvxf curl-7.86.0.tar.gz
   cd curl-7.86.0/
   ```

   b. *Compile the source code*

   ```
   mkdir curl_install
   ./configure --without-ssl --prefix=/path/to/new/curl/installation/curl-7.86.0/curl_install
   make
   make install
   ```

   c. *Add the curl installation directory to your PATH*

   ```
   export PATH=/path/to/new/curl/installation/curl-7.86.0/curl_install/bin:$PATH
   ```

ii. Install QUAST (Mikheenko et al., 2018)
   a. *Go to https://github.com/ablab/quast/releases, download the latest release and extract the tarball. Version 5.2.0 is shown as an example in the following commands.*

   ```
   cd /path/to/new/quast/installation
   curl -L --output quast-5.2.0.tar.gz https://github.com/ablab/quast/releases/download/quast_5.2.0/quast-5.2.0.tar.gz
   tar xvzf quast-5.2.0.tar.gz
   cd quast-5.2.0/
   ./install.sh
   ```

   b. *Add the QUAST installation directory to your PATH*

   ```
   export PATH=/path/to/new/quast/installation/quast-5.2.0:$PATH
   ```

iii. Install the SRA toolkit
   a. *Go to https://github.com/ncbi/sra-tools/wiki/01.-Downloading-SRA-Toolkit, and find the pre-built binary appropriate for your system. Download the archive and extract it. We use the centOS linux release v3.0.2 as an example below.*

   ```
   cd /path/to/new/sratools/installation
   curl -L --output sratoolkit.3.0.2-centos_linux64.tar.gz https://ftp-trace.ncbi.nlm.nih.gov/sra/sdk/3.0.2/sratoolkit.3.0.2-centos_linux64.tar.gz
   tar xvzf sratoolkit.3.0.2-centos_linux64.tar.gz
   ```

b. *Add the sratoolkit installation directory to your PATH*

```
export PATH=/path/to/new/sratools/installation/sratoolkit.3.0.2-
centos_linux64/bin:$PATH
```

3. Navigate to the directory where you want to run the ntLink tests, and download the sample long-read data.

```
cd /path/to/ntlink/test
fasterq-dump SRR10028109
```

*Once the command has finished, the reads will be available in the file* `SRR10028109.fastq`. *These reads are ~93-fold coverage C. elegans Oxford Nanopore long reads.*

4. Download the sample draft long-read assembly.
*This is a Flye (Kolmogorov et al., 2019) assembly of the long reads downloaded in the previous step.*

```
curl -L --output celegans_flye.fa
https://zenodo.org/record/7526395/files/celegans_flye.fa
```

5. Download a reference genome assembly for the *C. elegans* Bristol N2 strain.
*This assembly will be used in a later step when assessing the final assembly scaffolds using QUAST.*

```
curl -L --output celegans_reference.fa.gz
https://ftp.ncbi.nlm.nih.gov/genomes/all/GCA/000/002/985/GCA_000002985.3_
WBcel235/GCA_000002985.3_WBcel235_genomic.fna.gz
```

6. Run ntLink using the `ntLink` Makefile
*The specified values of k, w and overlap are the default values, but are included in the command to demonstrate how to set these parameters using the* `ntLink` *Makefile.*

```
ntLink scaffold target=celegans_flye.fa reads=SRR10028109.fastq k=32
w=100 t=5 overlap=True
```

7. Check the logs and output files to ensure that the run executed successfully.
*If ntLink completed successfully, the ntLink log should contain the message "Done ntLink! Final post-ntLink scaffolds can be found in: celegans_flye.fa.k32.w100.z1000.ntLink.scaffolds.fa". In addition, the final output scaffolds file "celegans_flye.fa.k32.w100.z1000.ntLink.scaffolds.fa" should be in the current working directory.*

8. Assess the final output scaffolds using abyss-fac (*de novo* approach) and QUAST (reference-based approach). See Table 2 for the expected statistics generated from these steps.
    a. Run abyss-fac using the input draft genome assembly and the post-ntLink genome assembly.

```
abyss-fac -G100e6 --count-ambig celegans_flye.fa
celegans_flye.fa.k32.w100.z1000.ntLink.scaffolds.fa
```

*The "`-G`" option specifies the genome size, which is approximately 100 Mbp for C. elegans. The "`--count-ambig`" option counts any ambiguous bases (ex. "N"s) in the output statistics.*
*After running this command, you will see that the NG50 length (at least half of the genome is in pieces at least this length) increases after ntLink scaffolding, and the number of sequences ('n') decreases.*

   b. Run QUAST to assess the input draft genome assembly and the post-ntLink genome assembly using the previously downloaded *C. elegans* reference assembly.

```
quast -t 5 -o quast_ntlink_bp1 -r celegans_reference.fa.gz --fast
--large celegans_flye.fa
celegans_flye.fa.k32.w100.z1000.ntLink.scaffolds.fa
```

*After running QUAST, all output files will be written to the* `quast_ntlink_bp1` *directory. The main statistics will be written to* `report.tsv`. *You will see that certain statistics, such as "# contigs (>= 0 bp)" and "NG50" will be the same as calculated by* `abyss-fac`. *Important reference-based statistics to look at include the "NGA50" and "# misassemblies". The NGA50 length is similar to the NG50 length, except that it uses alignment blocks for the calculation instead of sequence lengths. Therefore, it summarizes both the contiguity and correctness of the assemblies. QUAST defines a misassembly as a large-scale error in the* *assembly compared to the reference. Therefore, after scaffolding, we want to minimize the number of contigs and misassemblies, while maximizing the NG50 length and NGA50 length. Note that the QUAST executable will be* `quast.py` *if the tool was installed manually.*

**Table 2.** Expected results from scaffolding the *C. elegans* Flye assembly using ntLink with the steps documented in Basic Protocol 1. Compared to the baseline, scaffolding using ntLink with overlap detection increases the assembly NGA50 length approximately 1.7-fold, while also reducing the number of misassemblies.

| Assembly | Number of sequences >= 3 kbp | NG50 length (Mbp) | NGA50 length (Mbp) | Number of misassemblies |
| --- | --- | --- | --- | --- |
| Baseline assembly | 63 | 3.6 | 2.3 | 75 |
| Baseline assembly + ntLink | 33 | 6.8 | 3.7 | 66 |

## Basic Protocol 2: ntLink scaffolding with gap-filling

Basic Protocol 2 describes how to run ntLink scaffolding with a gap-filling step. In this protocol, instead of simply introducing ambiguous bases, ntLink fills gaps with sequence from the input long-read sequencing data. The initial steps of ntLink are executed as described in Basic Protocol 1. Then, an additional step is performed which fills-in the ntLink-induced scaffold gaps with bases from a representative read that supports the given join (the read that has the highest average number of minimizers in common with the incident contigs). Following this gap-filling step, the scaffolds are output in FASTA format. Because the gaps are filled with raw long-read sequence, we recommend polishing the output assembly using a long-read polishing tool such as ntEdit+Sealer (J. X. Li et al., 2022; Wong et al., 2022), Racon (Vaser et al., 2017) or Medaka (*Medaka: Sequence Correction Provided by ONT Research*, n.d.) following the ntLink scaffolding and gap-filling.

### Necessary Resources:

*Hardware*

This protocol requires a 64-bit Linux or MacOS operating system with sufficient RAM and available disk space (See Strategic Planning for more information).

*Software*

The following software must be installed and available in your PATH environment variable:
- SRA toolkit (v3.0.0+): (*https://github.com/ncbi/sra-tools*)
- curl: (*https://curl.se/*)
- Python 3.7+: (*https://www.python.org/*)
- ntLink (v1.3.7+): (*https://github.com/bcgsc/ntLink*)
- ABySS (v2.3.0+): (*https://github.com/bcgsc/abyss*)
- minimap2: (*https://github.com/lh3/minimap2*)
- Racon (v1.5.0+): (*https://github.com/lbcb-sci/racon*)
- QUAST (v5.2.0+): (*https://github.com/ablab/quast*)

*Files*

The input files for ntLink are long genome sequencing reads and a draft genome assembly. The long sequencing reads can be provided in FASTA or FASTQ format, either compressed with gzip or uncompressed. The input draft assembly to be scaffolded should be in FASTA format (multi-line or single-line).

*Sample Files*

The sample files used for this protocol are the same as used in Basic Protocol 1.

### Protocol steps:

1. Install the required software.

    *For more information about installing ntLink and all other dependencies other than minimap2 and Racon, please see the detailed instructions in Basic Protocol 1, steps 1-2.*

    *Installing protocol-specific dependencies minimap2 and Racon:*

*Option A: Using conda package manager*

i. If ntLink was installed using Option A of the Support Protocol, minimap2 and Racon can be installed in the same environment.

```
conda activate ntlink_env
conda install -y -c bioconda -c conda-forge minimap2 racon
```

*Option B: Installing from source*

i. Install minimap2
   a. For Linux, minimap2 provides pre-compiled binaries. Go to *https://github.com/lh3/minimap2/releases* to find the most recent pre-compiled binary. Here, we show downloading the v2.24 binary as an example:

   ```
   curl -L
   https://github.com/lh3/minimap2/releases/download/v2.24/minimap2-
   2.24_x64-linux.tar.bz2 | tar -jxvf -
   ```

   b. For MacOS, review the minimap2 dependencies (*https://github.com/lh3/minimap2*), download the most recent release tarball from *https://github.com/lh3/minimap2*, extract it and compile the code. We show downloading and compiling the v2.24 release as an example.

   ```
   curl -L --output minimap2-2.24.tar.bz2
   https://github.com/lh3/minimap2/releases/download/v2.24/minimap2-
   2.24.tar.bz2
   tar -jxvf minimap2-2.24.tar.bz2
   cd minimap2-2.24
   make
   ```

   c. Append the path to the minimap2 installation to your PATH environment variable.

   ```
   export PATH=/path/to/minimap2/installation:$PATH
   ```

ii. Install Racon (See *https://github.com/lbcb-sci/racon* for information about dependencies), and add the path to the Racon installation to your PATH environment variable.

```
git clone https://github.com/lbcb-sci/racon && cd racon
mkdir build && cd build
cmake -DCMAKE_BUILD_TYPE=Release ..
make
export PATH=/path/to/racon/install/build/bin:$PATH
```

2. Download the sample data. As the sample data for this protocol is the same as used for Basic Protocol 1, please see steps 3-5 of Basic Protocol 1 for full details about downloading the long reads, draft genome assembly and reference genome.

3. Change to the directory with the downloaded data, and run ntLink with the gap-filling option specified. The ntLink steps are powered by the `ntLink` Makefile.

   ```
   cd /path/to/ntlink/test
   ntLink scaffold gap_fill target=celegans_flye.fa reads=SRR10028109.fastq
   k=32 w=100 t=5
   ```

*Note that the k and w values specified are the default values, but are included in the command to illustrate how they can be set when running ntLink. The target* `gap_fill` *being specified in the command triggers the gap-filling stage after the initial scaffolding steps of ntLink.*

4. Check the logs and output files to ensure that the run executed successfully.

*If ntLink completed successfully, this message will be found in the logs: "Done ntLink! Final post-ntLink and gap-filled scaffolds can be found in: celegans_flye.fa.k32.w100.z1000.ntLink.scaffolds.fa". In addition, the final output scaffolds file "celegans_flye.fa.k32.w100.z1000.ntLink.scaffolds.fa" will be in the current working directory. The intermediate scaffold file before gap-filling is "celegans_flye.fa.k32.w100.z1000.stitch.abyss-scaffold.fa".*

5. Polish the gap-filled ntLink scaffolds. For illustrative purposes, we demonstrate polishing using Racon, but any long-read polishing tool can be utilized. This is an optional step in the pipeline, and can be bypassed if the integration of raw long reads (with a lower base quality) into the draft assembly is not a concern.

   a. *First, align the long reads to the draft assembly, and output the alignments in SAM format.*

   ```
   minimap2 -a -t 5 -x map-ont -o
   celegans_flye.fa.k32.w100.z1000.ntLink.scaffolds.SRR10028109.sam
   celegans_flye.fa.k32.w100.z1000.ntLink.scaffolds.fa
   SRR10028109.fastq
   ```

   b. *Next, run Racon, supplying the SAM file generated in step 5a.*

   ```
   racon -u -t 5 SRR10028109.fastq
   celegans_flye.fa.k32.w100.z1000.ntLink.scaffolds.SRR10028109.sam
   celegans_flye.fa.k32.w100.z1000.ntLink.scaffolds.fa 1>
   celegans_flye.fa.k32.w100.z1000.ntLink.scaffolds.racon-polished.fa
   ```

   c. Check the Racon output files to ensure that the run executed successfully. The final, polished assembly will be in the file "celegans_flye.fa.k32.w100.z1000.ntLink.scaffolds.racon-polished.fa", and the final log message in a successful Racon run will include "[racon::Polisher::] total =", along with the runtime.

6. Assess the final, polished output scaffolds using abyss-fac (reference-free) and QUAST (reference-based). See Table 3 for the expected results.

   a. *Run abyss-fac using the input draft genome assembly, the ntLink intermediate scaffolds file before gap-filling, and the final output scaffolds file after gap-filling and polishing.*

   ```
   abyss-fac -G100e6 --count-ambig celegans_flye.fa
   celegans_flye.fa.k32.w100.z1000.stitch.abyss-scaffold.fa
   celegans_flye.fa.k32.w100.z1000.ntLink.scaffolds.racon-polished.fa
   ```

   *See Basic Protocol 1, Step 8a for detailed information describing the abyss-fac output.*

   b. *Run QUAST to assess the input draft genome assembly, the ntLink intermediate scaffolds file before gap-filling, the output scaffolds after gap-filling and the final output scaffolds file after gap-filling and polishing.*

   ```
   quast -t 5 -o quast_ntlink_bp2 -r celegans_reference.fa.gz --fast
   --large --split-scaffold celegans_flye.fa
   celegans_flye.fa.k32.w100.z1000.stitch.abyss-scaffold.fa
   celegans_flye.fa.k32.w100.z1000.ntLink.scaffolds.fa
   celegans_flye.fa.k32.w100.z1000.ntLink.scaffolds.racon-polished.fa
   ```

   *See Basic Protocol 1, Step 8b for detailed information about the QUAST output. In addition to the QUAST statistics described in the previous protocol, for this protocol we are also interested in distinguishing between "Scaffold NG50/NGA50" and "Contig NG50/NGA50", which are available from the QUAST output in the* `quast_ntlink_bp2` *directory. The "Scaffold NG50" defines the sequence length where at least half of the genome is in sequences at least the NG50 length, with the "Scaffold NGA50" being the equivalent statistic, but calculated using alignment block lengths instead of sequence lengths. The "Contig NG50/NGA50" statistics are similar, except that the sequences are broken at ambiguous codes ("N"s) prior to the calculation. When the* `--split-scaffold` *option is specified for QUAST, it will output the statistics for the full input assembly ("Scaffold" statistics), and the assembly after breaking the sequences at regions of >= 10 Ns ("Contig" statistics, "_broken" added to filename). Therefore, the "Contig NG50/NGA50" statistics are a measure of contiguity as well as the number and distribution of gaps in the assembly. Furthermore, the QUAST statistic "# N's per 100 kbp" gives a direct measure of the number of ambiguous bases in the assembly. With efficient gap-filling, the "Contig" statistics will become closer to the "Scaffold" statistics, and the "# N's per 100 kbp" will decrease.*

**Table 3.** Expected results from scaffolding the *C. elegans* Flye assembly using ntLink with gap-filling based on the steps documented in Basic Protocol 2. Using the gap-filling feature of ntLink in addition to scaffolding increased the Contig NG50 and NGA50 lengths to be equivalent with their "Scaffold" statistic counterparts. Furthermore, while the intermediate ntLink scaffolds file prior to gap-filling had a sharp increase in the number of N's per 100 kbp (14.4 compared to 0.0 in the baseline), the gap-filling step sealed the majority of the gaps (14.4 vs. 0.02 N's per 100 kbp before and after gap-filling, respectively).

| Assembly | Number of sequences >= 3 kbp | Scaffold NG50 length (Mbp) | Contig NG50 length (Mbp) | Scaffold NGA50 length (Mbp) | Contig NGA50 length (Mbp) | Number of misassemblies | # N's per 100 kbp |
|---|---|---|---|---|---|---|---|
| Flye baseline | 63 | 3.6 | 3.6 | 2.3 | 2.3 | 75 | 0.00 |
| Flye + ntLink (before gap-filling) | 33 | 6.8 | 4.0 | 3.7 | 2.5 | 66 | 14.36 |
| Flye + ntLink | 33 | 6.8 | 6.8 | 3.9 | 3.9 | 64 | 0.02 |
| Flye + ntLink + Racon | 33 | 6.9 | 6.9 | 3.0 | 3.0 | 58 | 0.01 |

## Basic Protocol 3: Running in-code iterations of ntLink scaffolding

Basic Protocol 3 describes how to run multiple iterations, or rounds, of ntLink using a liftover-based approach. When scaffolding assemblies, the goal is to achieve the highest possible contiguity without sacrificing the correctness of the assembly. While running a single round of ntLink, as described in Basic Protocols 1 and 2, is very effective in improving upon a draft genome assembly from any technology, further gains are possible with additional rounds of ntLink. Using the in-code round capability of ntLink allows a user to maximize the contiguity of the final assembly without needing to manually run ntLink multiple times. To avoid re-mapping the reads at each round, ntLink lifts over the mapping coordinates from the input draft assembly to the output post-ntLink scaffolds, which can then be used for the next round of ntLink. The same process can be repeated as many times as needed, thus enabling multiple rounds of ntLink to be powered by a single instance of long-read mapping.

### Necessary Resources:

*Hardware*

This protocol requires a 64-bit Linux or MacOS operating system with sufficient RAM and available disk space (See Strategic Planning for more information).

*Software*

The following software must be installed and available in your PATH environment variable:
- SRA toolkit (v3.0.0+): (*https://github.com/ncbi/sra-tools*)
- curl: (*https://curl.se/*)
- Python 3.7+: (*https://www.python.org/*)
- ntLink (v1.3.7+): (*https://github.com/bcgsc/ntLink*)
- ABySS (v2.3.0+): (*https://github.com/bcgsc/abyss*)
- QUAST (v5.2.0+): (*https://github.com/ablab/quast*)

*Files*

The input files for ntLink are long genome sequencing reads and a draft genome assembly. The long sequencing reads can be provided in FASTA or FASTQ format, either compressed with gzip or uncompressed. The input draft assembly to be scaffolded should be in FASTA format (multi-line or single-line).

*Sample Files*

The sample files used for this protocol are the same as used in Basic Protocol 1.

### Protocol steps:

1. Install the required software.

   *For more information about installing ntLink and all other dependencies, please see detailed instructions in Basic Protocol 1, steps 1-2.*

2. Download the sample data. The sample data is the same as for Basic Protocols 1 and 2.

*For detailed instructions describing downloading the sample data, see Basic Protocol 1, steps 3-5.*

3. Run 3 rounds of ntLink scaffolding.

   *Change into a new directory, and create soft links so that the input files are accessible in the current working directory*

   ```
   cd /path/to/ntLink/test
   mkdir -p run_rounds && cd run_rounds
   ln -s ../celegans_flye.fa && ln -s ../SRR10028109.fastq
   ln -s ../celegans_reference.fa.gz
   ```

   *Option A: Run rounds of ntLink without gap-filling*

   ```
   ntLink_rounds run_rounds target=celegans_flye.fa reads=SRR10028109.fastq k=32 w=100 t=5 rounds=3 dev=True
   ```

   *Option B: Run rounds of ntLink with gap-filling*

   ```
   ntLink_rounds run_rounds_gaps target=celegans_flye.fa reads=SRR10028109.fastq k=32 w=100 t=5 rounds=3 dev=True
   ```

   *The dev=True option will retain all intermediate files. Although this is useful to be able to see all the file types generated by ntLink for working through this protocol, this option can be omitted for most runs. When omitted, some intermediate files will be automatically deleted to save disk space.*

4. Check the logs and output files to ensure that the ntLink run executed successfully.

   *After the ntLink command has completed, check the log for this final message, which indicates a successful run: "Done ntLink rounds! Final scaffolds found in celegans_flye.fa.k32.w100.z1000.ntLink.3rounds.fa". This message also indicates the FASTA file which contains the final, scaffolded assembly sequences.*

5. Use abyss-fac (*de novo* approach) and QUAST (reference-based approach) to assess the genome assembly after each round of ntLink scaffolding, and compare the results to the initial baseline assembly. See Figure 2 for a summary of the expected assembly statistics.

   a. Reference-free analysis of the ntLink output scaffolds using abyss-fac.

   *If Option A was followed in Step 3:*

   ```
   abyss-fac --count-ambig -G100e6 celegans_flye.fa
   celegans_flye.fa.k32.w100.z1000.ntLink.fa
   celegans_flye.fa.k32.w100.z1000.ntLink.ntLink.fa
   celegans_flye.fa.k32.w100.z1000.ntLink.3rounds.fa
   ```

   *If Option B was followed in Step 3:*

   ```
   abyss-fac --count-ambig -G100e6 celegans_flye.fa
   celegans_flye.fa.k32.w100.z1000.ntLink.gap_fill.fa
   celegans_flye.fa.k32.w100.z1000.ntLink.ntLink.gap_fill.fa
   celegans_flye.fa.k32.w100.z1000.ntLink.3rounds.fa
   ```

   *See Basic Protocol 1, Step 8a for detailed information describing the abyss-fac output.*

   b. Reference-based analysis of the ntLink output scaffolds using QUAST.

   *If Option A was followed in Step 3:*

   ```
   quast -t 5 -o quast_ntlink_bp3 -r celegans_reference.fa.gz --fast
   --large --split-scaffold celegans_flye.fa
   celegans_flye.fa.k32.w100.z1000.ntLink.fa
   celegans_flye.fa.k32.w100.z1000.ntLink.ntLink.fa
   celegans_flye.fa.k32.w100.z1000.ntLink.3rounds.fa
   ```

   *If Option B was followed in Step 3:*

   ```
   quast -t 5 -o quast_ntlink_bp3 -r celegans_reference.fa.gz --fast
   --large --split-scaffold celegans_flye.fa
   celegans_flye.fa.k32.w100.z1000.ntLink.gap_fill.fa
   celegans_flye.fa.k32.w100.z1000.ntLink.ntLink.gap_fill.fa
   celegans_flye.fa.k32.w100.z1000.ntLink.3rounds.fa
   ```

*See Basic Protocol 1, Step 8b and Basic Protocol 2, Step 6b for detailed information about the QUAST output. The QUAST output will be written to the directory named* `quast_ntlink_bp3`. *Note that if QUAST was installed from source, the executable will be named* `quast.py`.

*Note that the final scaffolds file name will be the same whether Option A or Option B was followed, but the names of the files from intermediate rounds differ slightly.*

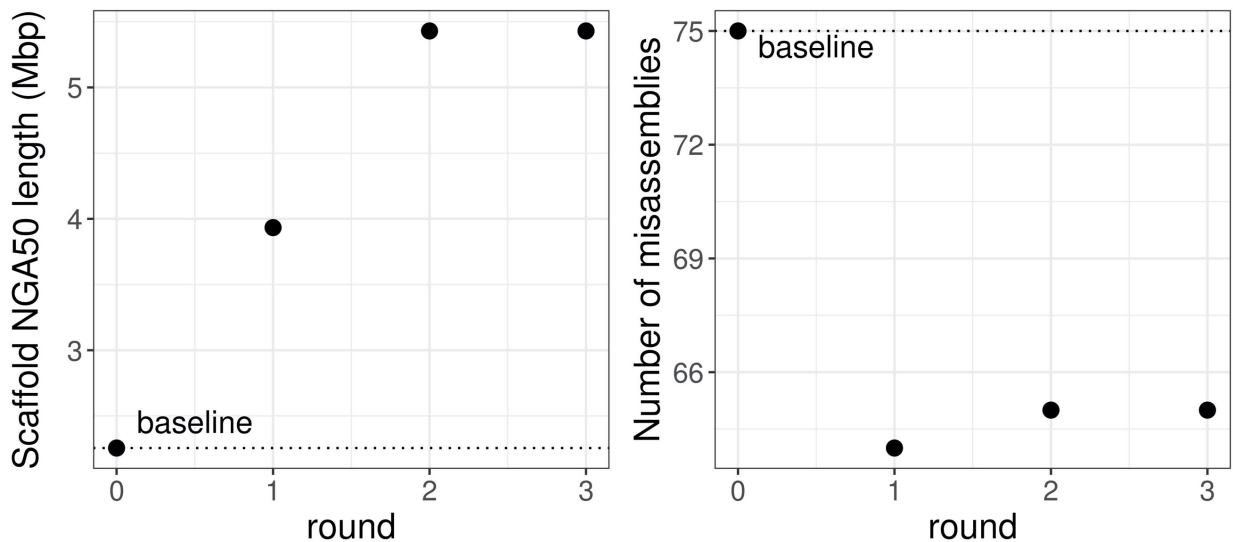

**Figure 2.** Results from analyzing the output files from Basic Protocol 3 (ntLink Option B) using QUAST. Running additional rounds of ntLink results in a more contiguous final assembly (Scaffold NGA50 length of 3.9 Mbp and 5.4 Mbp for round 1 and 2, respectively). The third round of scaffolding did not produce any additional joins, showing that further ntLink rounds would not benefit the assembly. More rounds may be required when the baseline assembly is more fragmented, or a larger genome is being scaffolded.

## Alternate Protocol 1: Generating long-read to contig mappings with ntLink

Although ntLink is most commonly used as a scaffolding tool, the minimizer-based mapping functionality that enables assembly scaffolding can also be run in isolation. In any mode, ntLink first maps the input long reads to the input draft genome. When using ntLink in the default scaffolding mode, these mappings are parsed to infer evidence that supports ordering and orienting contigs into scaffolds. However, this mapping information can also be simply output to a file in a standard format and used for other downstream applications. In Alternate Protocol 1, we demonstrate the mapping mode of ntLink, which outputs the mappings in a standard (Pairwise mApping Format) PAF format.

**Necessary Resources:**

*Hardware*

This protocol requires a 64-bit Linux or MacOS operating system with sufficient RAM and available disk space (See Strategic Planning for more information).

*Software*

The following software must be installed and available in your PATH:
- SRA toolkit (v3.0.0+): (*https://github.com/ncbi/sra-tools*)
- curl: (*https://curl.se/*)
- Python 3.7+: (*https://www.python.org/*)
- ntLink (v1.3.7+): (*https://github.com/bcgsc/ntLink*)
- miller: (*https://miller.readthedocs.io/en/latest/*)

*Files*

The input files for mapping with ntLink are long genome sequencing reads and a draft genome assembly. The long sequencing reads can be provided in FASTA or FASTQ format, either compressed with gzip or uncompressed. The input draft assembly to be scaffolded should be in FASTA format (multi-line or single-line).

*Sample Files*

In this protocol, we will map the same Oxford Nanopore *C. elegans* long reads as used in the basic protocols to a short-read *C. elegans* AbySS (Jackman et al., 2017) assembly. The AbySS assembly is available from https://zenodo.org/record/7526395/files/celegans_abyss.fa.

## Protocol steps:

1. Install the required software.

   *For more information about installing ntLink and other dependencies, please see detailed instructions in Basic Protocol 1, steps 1-2.*

2. Install the protocol-specific dependency, miller.

   *Option A: Use conda package manager*

   i. If Option A of the Support Protocol was used to install ntLink, miller can be installed in the same conda environment.

   ```
   conda activate ntlink_env
   conda install -y -c conda-forge miller
   ```

   *Option B: Install from source*

   i. Go to the miller releases page (*https://github.com/johnkerl/miller/releases*) and find the pre-built binary that is appropriate for your system.

   ii. Download the binary from the releases page, and extract the compressed tarball. A Linux pre-built binary is shown for illustration purposes, but the path to any pre-built miller binary can be used in this step.

   ```
   curl -L --output miller_download.tar.gz
   https://github.com/johnkerl/miller/releases/download/v6.5.0/miller-6.5.0-
   linux-amd64.tar.gz
   tar xvzf miller_download.tar.gz
   ```

   iii. Append the path to the miller installation to your PATH environment variable.

   ```
   export PATH=/path/to/miller/installation/miller/:$PATH
   ```

3. Download the sample long reads. This file is the same as used for Basic Protocols 1, 2 and 3.

   *For detailed instructions describing downloading the sample long reads, see Basic Protocol 1, step 3.*

4. Download the sample draft AbySS (Jackman et al., 2017) short-read assembly.

   ```
   curl -L --output celegans-abyss.fa
   https://zenodo.org/record/7526395/files/celegans_abyss.fa
   ```

5. Run ntLink to map the sample long reads to the draft assembly. Ensure that the downloaded sample data files are in your current working directory.

   ```
   ntLink pair target=celegans-abyss.fa reads=SRR10028109.fastq t=5
   sensitive=True paf=True
   ```

6. Check the logs and output files to ensure that the run executed successfully.

   *If the ntLink mapping completed successfully, the log messages from ntLink should finish with a time stamp and "DONE!". Furthermore, a PAF-formatted mapping file called "celegans-abyss.fa.k32.w100.z1000.paf" should be in your current working directory.*

7. Assess the mapping file output from ntLink. See Table 4 for the expected mapping statistics.
   a. To make the generation of summary statistics more straightforward with miller, add column labels to the PAF file.

```
cat celegans-abyss.fa.k32.w100.z1000.paf | mlr --tsv --implicit-
csv-header label
read,read_len,read_start,read_end,strand,contig,contig_len,contig_
start,contig_end,num_minimizers,len_mapping,mapping_qual >
celegans-abyss.fa.paf.mlr.tsv
```

  b. Count the total number of mappings of the long reads to the query contigs.

```
mlr --tsv stats1 -a count -f read celegans-abyss.fa.paf.mlr.tsv
```

  c. Calculate the average mapping block length.

```
cat celegans-abyss.fa.paf.mlr.tsv |mlr --tsv stats1 -a mean -f
len_mapping
```

  d. Calculate the average number of read mappings per draft assembly contig.

```
cat celegans-abyss.fa.paf.mlr.tsv |mlr --tsv cut -f read,contig
then uniq -g read,contig then stats1 -a count -f contig -g contig
then stats1 -a mean -f contig_count
```

**Table 4**. Expected results from assessing the ntLink mappings of the sample *C. elegans* long reads to the draft ABySS assembly in Alternate Protocol 1.

| | |
|---|---|
| **Total number of read mappings** | 622,975 |
| **Average mapping block length (bp)** | 5,647.7 |
| **Average number of distinct mapped reads per contig** | 105.2 |

## Alternate Protocol 2: Using ntLink mappings for genome assembly correction with Tigmint-long

As described in Alternate Protocol 1, the mapping functionality in ntLink can be used to inform scaffolding, the most common use of ntLink, or separately to provide mapping information that can be used by other downstream pipelines. One such alternate application is Tigmint-long (Coombe et al., 2021), a *de novo* genome assembly correction tool which utilizes information in long reads to detect and cut at putative misassemblies. In the default mode, Tigmint-long simulates pseudo-linked reads from the long reads. This involves breaking the long reads into tiles, which represent short-read fragments, then generating paired-end reads from the fragments. Each read pair from the same long read is assigned the same barcode, adhering to the expected format for linked reads. These reads are then mapped to the draft assembly using minimap2 (H. Li, 2018), and these mappings are parsed to look for regions of the draft assembly that are not well-supported by the reads. However, as only approximate mappings are required, ntLink mapping can be used in place of minimap2. As ntLink uses more streamlined mapping logic, the reads do not need to be broken into pseudo-linked reads prior to mapping, thus eliminating a step in the pipeline. The output of Tigmint-long is a contigs file in FASTA format, where the sequences are broken at putative misassemblies.

*Hardware*

This protocol requires a 64-bit Linux or MacOS operating system with sufficient RAM and available disk space (See Strategic Planning for more detail).

*Software*

The following software must be installed and available in your PATH:
- SRA toolkit (v3.0.0+): (*https://github.com/ncbi/sra-tools*)
- curl: (*https://curl.se/*)
- Python 3.7+: (*https://www.python.org/*)
- ntLink (v1.3.7+): (*https://github.com/bcgsc/ntLink*)
- Tigmint (v1.2.9+): (*https://github.com/bcgsc/tigmint*)
- QUAST (v5.2.0+): (*https://github.com/ablab/quast*)

*Files*

The input files for ntLink are long genome sequencing reads and a draft genome assembly. The long sequencing reads can be provided in FASTA or FASTQ format, either compressed with gzip or uncompressed. The input draft assembly to be corrected should be in FASTA format (multi-line or single-line).

## Protocol steps:

1. Install the required software used in the earlier protocols, if not already installed.

   *For more information about installing ntLink and other common dependencies, please see detailed instructions in Basic Protocol 1, steps 1-2.*

2. Install the protocol-specific dependency Tigmint.

   *Option A: Use conda package manager*

   i.  If Option A of the Support Protocol was used to install ntLink, Tigmint can be installed in the same conda environment.

   ```
   conda activate ntlink_env
   conda install -y -c bioconda -c conda-forge tigmint 'samtools>=1.10'
   ```

   *Option B: Install from source*

   i.  Consult the README in the Tigmint GitHub repository (*https://github.com/bcgsc/tigmint*) to ensure that the required dependencies are installed.

   ii. Go to the releases page for Tigmint (*https://github.com/bcgsc/tigmint/releases*) and find the most recent release tarball. Download and extract this tarball in the directory where you would like Tigmint to be installed. To demonstrate the required commands, Tigmint v1.2.9 is shown below, but the URL can be substituted for any later release of Tigmint.

   ```
   curl -L --output tigmint-1.2.9.tar.gz https://github.com/bcgsc/tigmint/releases/download/v1.2.9/tigmint-1.2.9.tar.gz
   tar xvzf tigmint-1.2.9.tar.gz
   cd tigmint-1.2.9
   ```

   iii. Compile the required binaries

   ```
   cd src
   make
   ```

   iv. Append the path to the Tigmint installation to your PATH environment variable.

   ```
   export PATH=/path/to/tigmint/install/tigmint-1.2.9/bin:$PATH
   ```

3. Download the sample *C. elegans* long reads and reference genome. These files are the same as used in Basic Protocols 1-3.

   *For detailed instructions describing downloading this sample data, see Basic Protocol 1, steps 3 and 5.*

4. Download the sample draft *C. elegans* ABySS short-read assembly. This draft assembly FASTA is the same as used in Alternate Protocol 1.

   *For detailed instructions describing downloading the ABySS short-read assembly, see Alternate Protocol 1, step 4.*

5. Run Tigmint-long on the draft *C. elegans* ABySS assembly using ntLink mapping and the downloaded long reads to detect and cut at putative misassemblies. Ensure that the downloaded sample data files are in your current working directory.

   ```
   tigmint-make tigmint-long draft=celegans-abyss reads=SRR10028109 mapping=ntLink t=5 span=2
   ```

6. Check the logs and output files from Tigmint-long to ensure that the run executed successfully.

   *A successful run of Tigmint will finish with the following log messages: "Cutting assembly at breakpoints… DONE!". There will also be a file named "celegans-abyss.cut500.tigmint.fa" in your current working directory which contains the corrected draft assembly sequences.*

7. Use the reference-based assessment tool QUAST to compare the contiguity and correctness of the corrected genome assembly and the initial baseline assembly. See Table 5 for results from running QUAST on these assemblies.

    ```
    quast -t 5 -o quast_tigmint_ap2 -r celegans_reference.fa.gz --fast --large celegans-abyss.fa celegans-abyss.cut500.tigmint.fa
    ```

    *See Basic Protocol 1, Step 8b for detailed information about the QUAST output. Note that if QUAST was installed from source, the executable will be named* `quast.py`. *The QUAST results will be written to the* `quast_tigmint_ap2` *directory.*

**Table 5.** Results from analyzing the baseline and Tigmint-long corrected *C. elegans* AbySS assembly. Following Tigmint-long misassembly correction using mappings from ntLink, the number of misassemblies decreases by more than 2-fold.

| Assembly | Number of sequences >= 3 kbp | Scaffold NG50 length (bp) | Scaffold NGA50 length (bp) | Number of misassemblies |
| --- | --- | --- | --- | --- |
| **ABySS baseline** | 4,552 | 30,293 | 27,015 | 433 |
| **ABySS + Tigmint-long** | 4,793 | 27,564 | 27,015 | 162 |

## Support Protocol: Installing ntLink

ntLink can be installed using the conda package manager or from the source code. For a more straightforward installation process, and to ensure that all dependencies are properly installed, we recommend installing ntLink using conda.

### Necessary Resources:

#### *Hardware*

ntLink requires a 64-bit Linux or MacOS operating system with sufficient RAM and available disk space (See Strategic Planning for more detail).

#### *Software*

Miniconda (https://docs.conda.io/en/latest/miniconda.html)

### Protocol steps:

**Option A: Installing ntLink using the conda package manager**

1. If miniconda is not already installed:
   i. Download the miniconda3 bash installer: https://docs.conda.io/en/latest/miniconda.html
   ii. Run the installer script:
      a. *On MacOS:*
         `bash Miniconda3-latest-MacOSX-x86_64.sh`
      b. *On Linux:*
         `bash Miniconda3-latest-Linux-x86_64.sh`
   iii. Follow the installer prompts
   iv. Close and re-open your terminal window to finalize the installation

2. Create a new conda environment

    `conda create -n ntlink_env`

3. Activate the new conda environment

    `conda activate ntlink_env`

4. Install ntLink in the environment

    `conda install -y -c bioconda -c conda-forge ntlink`

**Option B: Installing ntLink from the source code**

1. Install the following dependencies, and ensure that each is available in your PATH environment variable. We recommend installing the dependencies using a package manager such as conda. Otherwise, visit the tool homepages for information about installing from source.
    - Python 3.7+
    - Python modules:
        - Numpy: (https://numpy.org/)
        - Python-igraph: (https://igraph.org/python/)
    - btllib: (*https://github.com/bcgsc/btllib*)
    - ABySS (v2.3.0+): (*https://github.com/bcgsc/abyss*)
    - Zlib: (*https://zlib.net/*)
    - Make: (*https://www.gnu.org/software/make/*)

2. Change your directory to the desired folder for the ntLink installation, then clone the ntLink repository from GitHub.
    ```
    cd /path/to/desired/location/for/ntlink
    git clone https://github.com/bcgsc/ntLink.git
    ```

3. Append the location of the ntLink installation to your PATH environment variable
    ```
    export PATH=/path/to/desired/location/for/ntlink/ntLink:$PATH
    ```

**Checking your installation**

To verify that your installation is working properly, you can follow any of the basic protocols, or run the small demo provided on GitHub.

*Running test demo*

1. If you haven't already cloned ntLink during the installation process, clone the GitHub repository to download the small test demo.
    ```
    git clone https://github.com/bcgsc/ntLink.git
    ```

2. Change your working directory to the cloned ntLink repository, then to the directory containing the test demo script.
    ```
    cd ntLink/tests
    ```

3. Run the provided demo shell script
    ```
    ./test_installation.sh
    ```

4. If the test was successful, indicating that your installation is working as expected, you will see this message: "Done tests! Compare your generated files with the files in the expected_outputs folder to ensure the tests were successful.".

## Guidelines for understanding results

For all ntLink runs, it is important to look through the log messages to ensure that there are no errors. If there are error messages at any stage, the results are not reliable, and the error(s) needs to be resolved prior to any downstream genome analysis. See Table 6 for some common errors and suggested solutions.

Running ntLink for scaffolding or mapping will generate various intermediate files. For scaffolding runs, the most important output file is the FASTA file containing the final, improved scaffolds. However, the other intermediate files contain useful information about both the evidence used for generating the final scaffolds as well as the composition of the output scaffolds themselves. The constructed scaffold graph is output in DOT format (".scaffold.dot"). In this graph, the nodes are contigs, and the directed edges represent long-read evidence between the incident contigs. When running ntLink scaffolding, this graph is traversed using abyss-scaffold (Jackman et al., n.d.) to produce the final ordered and oriented scaffolds. The ntLink output files with the suffices "trimmed_scafs.path" and "trimmed_scafs.agp" each describe the composition of the output scaffolds in different formats. These files allow the user to deduce the order and orientation of the input contigs in the output scaffolds, as well as any gap sequences between the contigs. The ".path" format describes one scaffold per tab-separated line, with the first column denoting the scaffold name, and the second listing the order and orientation of the contigs, with gap sizes indicated by "<number>N". The ".agp" file follows the standard AGP specifications. The AGP file with the suffix "gap_fill.fa.agp" is only generated when the gap-filling step of ntLink is performed, and additionally reports the identity and coordinates of the input long reads used to fill gaps.

Following successful ntLink scaffolding, it is expected that there will be fewer sequences in the scaffolded assembly compared to the baseline assembly, since input contigs will be joined together to form scaffolds. Consequently, the contiguity of the scaffolded assembly (as assessed by abyss-fac, QUAST or other assembly assessment tools) is expected to increase. If there is no change in the contiguity or number of sequences, it is possible that parameters such as *k* and *w* (controlling the generation of the minimizers) need to be optimized (See Critical Parameters). For example, if using a long-read dataset with a high error rate, a smaller *k* value may be needed to increase the sensitivity of the long-read mapping. When running rounds of ntLink, it is expected that the contiguity will not increase after several rounds, as demonstrated in Basic Protocol 3. This is not a cause for concern, but just an indication that the long-read evidence leveraged by ntLink may be exhausted.

As demonstrated in the protocols, it is important to analyse the output scaffold FASTA files with tools such as abyss-fac or QUAST to assess the scaffolding success. While abyss-fac analysis does not require a reference, QUAST is reference-based, and is thus not suitable for all studies. For assembly projects without an available reference genome, or if many structural variants are expected, BUSCO (Manni, Berkeley, Seppey, Simão, et al., 2021; Manni, Berkeley, Seppey, & Zdobnov, 2021) is a useful tool for reference-free assessment of the assembly quality. BUSCO, or Benchmarking Universal Single Copy Orthologs, searches the input genome assembly for genes that are evolutionarily expected to be found in single copy. Since BUSCO assesses the assembly completeness in the gene space, it provides complementary information to the reference-free contiguity metrics.

Following ntLink scaffolding and quality control of the resulting assembly, there are a variety of downstream analyses that can be performed, from comparative genomics to annotation. The direction that these analyses take will be guided by the particular research lab and study focus, making this assembly stage broadly important.

## Commentary

### Background Information

Scaffolding tools, such as ntLink, can play important roles in *de novo* assembly pipelines through further improving upon draft assemblies. Multiple new features and modes have been integrated into ntLink to help users obtain the best possible assemblies and results from their sequencing data. The efficiency of the new and existing features of ntLink are largely attributable to the use of minimizer sketches for the various mapping tasks.

As described in Roberts et al. (2004) and implemented in btllib (Nikolić et al., 2022), ntLink generates ordered minimizer sketches by first breaking the input sequences into their constituent *k*-mers (substrings of length *k*), and computing a hash value for each *k*-mer using ntHash2 (Kazemi et al., 2022). Then, for each *w* (window size) *k*-mers, the *k*-mer with the smallest hash value is chosen as the minimizer for that window. Sliding this window across the entire sequence generates the ordered minimizer sketch, a particular subset of *k*-mers (represented by hash values) which is much smaller than the entire *k*-mer spectrum. Using this sketching approach for sequence mapping provides a great computational advantage for ntLink in both memory usage and time efficiency, enabling ntLink to scale to large genomes (Table 1).

The newly developed overlap detection, gap-filling and liftover-based round functionalities of ntLink benefit the final quality of the assemblies and allow the scaffolder to be more flexible to the specific needs of the users. Prior to the integration of overlap detection, ntLink would simply join sequences end-to-end with an intervening gap, whether the sequences had a putative overlap or not. This could lead to small insertion misassemblies being introduced at the join point, which could have a negative impact on such downstream applications as annotation, if the insertion is in a gene region, for example. The overlap detection feature resolves these overlaps, avoiding the introduction of small misassemblies at the join point and allowing for a cleaner join. When using ntLink to scaffold assemblies without the gap-filling feature, the regions between joined contigs that have a gap between them, or missing genomic sequence, are filled with ambiguous bases ("N"s). While it is valuable to have sequences ordered and oriented relative to one another, there is also genomic information in those gaps that will then be missing in the assembly. Gap-filling can be performed as a downstream, often computationally intensive, assembly step (Chu et al., 2019; Paulino et al., 2015), but performing gap-filling within ntLink is efficient and effective in recovering these missing regions. Finally, running liftover-based rounds of ntLink enables additional improvements to the draft assembly by fully leveraging the long-read evidence, while also avoiding the computational burden of re-mapping the reads at each round.

Other state-of-the-art long-read assemblers, such as LRScaf and OPERA-LG, rely on sequence alignments instead of mapping and do not provide users with the same features and flexibility as ntLink. Neither LRScaf nor OPERA-LG provides gap-filling functionality, nor an in-code approach for running rounds of scaffolding. Therefore, if a user wants to run multiple rounds of scaffolding, they would have to do so in a naïve manner (manually executing ntLink multiple times). Furthermore, while some long-read scaffolding tools such as LRScaf do also have logic to deal with overlapping joined sequences, their algorithms use alignments, while ntLink uses a more lightweight minimizer-mapping guided approach. OPERA-LG is not currently maintained (the last release was in 2016), so may not properly leverage more recent improvements in both sequencing and bioinformatics technologies.

Finally, we also demonstrate the flexibility of using the mapping functionality of ntLink for other applications in Alternate Protocols 1 and 2. Sequence mapping is a foundational process in bioinformatics, and often exact coordinates are not needed for the desired application. In this case, ntLink is a great resource for lightweight mapping, which can find numerous applications such as in misassembly correction (as demonstrated in Alternate Protocol 2), targeted assembly and targeted polishing, to name a few.

## Critical Parameters

### *k (k-mer size) and w (window size)*

The *k* and *w* parameters control the generation of minimizers for mapping the long reads to the draft assembly in ntLink, and are therefore the most influential parameters. Generally, the default settings (*k*=32, *w*=100) produce good results for a variety of input assemblies and reads, but in order to obtain the best final scaffolds, these parameters can be optimized using a grid search. If undertaking a grid search, the approximate recommended ranges of *k* and *w* to test would be *k*=[24-80] and *w*=[75-1000]. Generally, we recommend a lower *k* and *w* setting when the long reads and/or draft assembly are more erroneous. However, if the draft assembly is very contiguous and/or the base quality of the input data is high, higher values can be successful.

## Troubleshooting

The `ntLink` Makefile should complete with an exit code of 0 and a message indicating a successful run. If an error occurs, the pipeline should stop running and output an error message. Some common errors are documented in Table 6. If you encounter additional errors not discussed in Table 6, please create new issue at the ntLink GitHub repository (https://github.com/bcgsc/ntLink).

## Advanced Parameters

There are several ntLink parameters that can be tweaked in addition to *k* and *w* that may provide benefits for more advanced users. The default settings of each of these parameters have been found to work well for most assemblies.

### *z (minimum contig length)*

By default, only sequences greater than 1 kb (z=1000) will be considered for integration into an output scaffold. Depending on the contiguity of the input draft assembly, the user may want to adjust this parameter if the input assembly is very contiguous (increase z) or very fragmented (decrease z).

### *a (minimum number of anchoring reads)*

When ntLink parses the long-read evidence to create edges in the scaffold graph, it requires (by default) at least one 'anchoring read' for an edge to retained. An 'anchoring read' is defined as a read that has at least 2 mapped minimizers on each contig in the putative pair. If more stringent scaffold pairing is desired, this parameter can be increased to require more 'anchoring reads' before retaining an edge.

### *v (verbose benchmarking mode)*

If the user specifies `v=1` in their ntLink command, the time and peak memory will be tracked for each step, and output to separate files. This option can be useful when benchmarking the execution of ntLink.

### *soft_mask (soft mask filled gaps)*

If `soft_mask=True` is specified in the ntLink command when gap-filling is enabled, the gaps will be filled with lowercase bases instead of uppercase bases. This soft masking could be useful for downstream analyses such as targeted polishing, for example.

**Table 6.** Sources and Solutions to Potential Errors

| Problem | Possible Cause | Solution |
|---|---|---|
| Error "make: *** No rule to make target" | Input files are not in the current working directory, or full paths to files are used. | Make soft links to ensure that input files are available in the current working directory, and do not use absolute or relative paths to input files. |
| Error "zsh: no such option: pipefail" | An older version of zsh is installed, or zsh is missing. | Install zsh or update to the newest version. |
| "UnsatisfiableError" when installing ntLink using conda | Incompatible versions of tools were previously installed in the conda environment. | Install ntLink in a fresh conda environment. |
| "source file is in invalid format!" when running ntLink | Input long reads or draft assembly files are in an unexpected format. | Ensure that the input long reads are in correctly-formatted FASTA or FASTQ format (gzipped or uncompressed), and the draft assembly is in FASTA format (uncompressed). |
| "ModuleNotFoundError: No module named <package_name>" error | A required python module is not installed properly. | Ensure that the expected python version is being used (ex. with `which python3`), and install the missing package (using `conda` or `pip`). |
| "Error 127" in ntLink log file after `abyss-scaffold` step | ABySS is not installed, or not found in your PATH environment variable. | Install ABySS if needed, and ensure that the ABySS executables are found in your PATH. |
| Running `ntLink -h` prints the `make` help page instead of the ntLink help page | ntLink requires parameters to be specified in the form "variable_name=variable_value", so using options with the form "-<letter>" will specify options to `make` itself, not the ntLink program. | Run `ntLink help` to see the full ntLink help page. |
| Contiguity gains post-ntLink are minimal | Incorrect selection of k/w | If ntLink makes minimal joins, it is likely that the *k* and *w* parameters specified are not optimal. See the *Critical Parameters* section for more details about setting *k* and *w*. |
| Error when running Tigmint-long: "samtools: error while loading shared libraries" | An older version of samtools is installed. | Update the samtools installation. |

## Conflict of interest

The authors declare that they have no conflicts of interest.

## Data availability

The sample assemblies that support the protocol are available from https://doi.org/10.5281/zenodo.7526395. The sample long reads and reference genome are publicly available under SRA accession SRR10028109 and GenBank accession GCA_000002985.3, respectively.

## Acknowledgements

This study is supported by the Canadian Institutes of Health Research (CIHR) [PJT-183608] and the National Institutes of Health [2R01HG007182-04A1]. The content of this article is solely the responsibility of the authors, and does not necessarily represent the official views of the National Institutes of Health or other funding organizations. The funding organizations did not have a role in the design of the study, the collection, analysis and interpretation of the data, or in writing the manuscript.

# ntLink: a toolkit for *de novo* genome assembly scaffolding and mapping using long reads

Lauren Coombe, René L. Warren, Johnathan Wong, Vladimir Nikolic, Inanc Birol

## Supplementary Tables

**Supplementary Table 1. Accessions for the long-read datasets listed in Table 1.**

| Species | Fold coverage of long reads | Accession(s) |
|---|---|---|
| *Caenorhabditis elegans* | 93 | SRR10028109 |
| *Oryza sativa* | 62 | SRR10589512- SRR10589711 |
| *Solanum lycopersicum* | 72 | ERR6668574 (Subsampled to ~72-fold) |
| *Homo sapiens* | 50 | s3://ont-open-data/gm24385_2020.11/analysis/r9.4.1/20201026_1644_2-E5-H5_PAG07162_d7f262d5/guppy_v4.0.11_r9.4.1_hac_prom/basecalls.fastq.gz |

**Supplementary Table 2. Baseline assemblies used for the example ntLink runs listed in Table 1.**

| Species | Sequencing reads used for baseline assembly | Assembler used for baseline assembly | Baseline assembly parameters |
|---|---|---|---|
| *Caenorhabditis elegans* | Long reads (SRR10028109) | Flye (v2.9.1) | --nano-raw SRR10028109.fastq -g100m -t48 |
| *Oryza sativa* | Long reads (SRR10589512-SRR10589711) | GoldRush (v1.0.0) goldtigs* | G=373e6 P=10 a=1 m=20000 polisher_mapper=minimap2 span=2 dist=500 t=48 polisher=goldrush-edit |
| *Solanum lycopersicum* | Long reads (ERR6668574, subsampled to 72-fold) | GoldRush (v1.0.0) goldtigs* | G=824e6 P=15 a=1 m=20000 polisher_mapper=minimap2 span=2 dist=500 t=48 polisher=goldrush-edit |
| *Homo sapiens* | Short reads (SRR11321732) | ABySS (v2.2.3) | j=48 k=112 kc=3 B=150G l=40 s=1000 v=-v q=15 H=4 S=1000-10000 N=9 N=5-20 pelib1_de=-n5 |

*GoldRush goldtigs are the polished, corrected golden path sequences generated from the GoldRush *de novo* genome assembler.